\documentclass[fleqn,usenatbib]{mnras}

\usepackage{newtxtext,newtxmath}

\usepackage[T1]{fontenc}
\usepackage{ae,aecompl}

\usepackage{graphicx}	
\usepackage{amsmath}	
\usepackage{amssymb}	
\usepackage{amsmath}

\title[Microlensing Time delays]{Gravitational Microlensing Time Delays at High Optical Depth:  
Image Parities and the Temporal Properties of Fast Radio Bursts
}

\author[Geraint F. Lewis]{
Geraint F. Lewis\thanks{E-mail: geraint.lewis@sydney.edu.au}
\\
Sydney Institute for Astronomy, School of Physics, A28, The University of Sydney, NSW 2006, Australia
}

\date{Accepted XXX. Received YYY; in original form ZZZ}

\pubyear{2020}

\begin{document}
\label{firstpage}
\pagerange{\pageref{firstpage}--\pageref{lastpage}}
\maketitle

\begin{abstract}
Due to differing gravitational potentials and path lengths, gravitational lensing induces time delays between multiple images of a source which, 
for solar mass objects, is of order $\sim10^{-5}$ seconds. 
If an astrophysically compact source, such as a Fast Radio Burst (FRB), is observed through a region with a high optical depth of such microlensing masses, this gravitational lensing time delay can be imprinted on short timescale transient signals.     
In this paper, we consider the impact of the parity of the macroimage on the resultant microlensing time delays. It is found that this parity is directly imprinted on the microlensing signal, with macroimages formed at minima of the time arrival surface beginning with the most highly magnified microimages and then progressing to the fainter microimages. At macroimages at the maxima of the time arrival surface, this situation is reversed, with fainter images observed first and finishing with the brightest microimages. For macroimages at saddle-points, the signal again begins with fainter images, followed by brighter images before again fading through the fainter microimages.     
The growing populations of cosmologically distant bursty transient sources will undoubtedly result in the discovery of strong lensed, multiply imaged FRBs, which will be susceptible to microlensing by compact masses.
With the temporal resolution being offered my modern and future facilities, the detection of microlensing induced time delays will reveal the parities of the gravitational lens macroimages, providing  
additional constraints on macrolensing mass models and improving the efficacy of these transient sources as a cosmological probes.
\end{abstract}




\section{Introduction}
Gravitational lensing by a massive object can produce multiple images of a distant source \citep[e.g.][]{PhysRev.51.290,1992grle.book.....S}. Due to differing paths between the source and the observer, photons can travel different distances and experience different gravitational potentials, leading to a time delay between the images that can be used as a cosmological probe \citep[see][]{1964MNRAS.128..307R,1966MNRAS.132..101R,2019arXiv190704869W,2020MNRAS.492.1102N}. 

The presence of small-scale masses, such as stars, planets and black holes, within a galactic scale lens can result in additional multiple imaging of a distant source \citep{1979Natur.282..561C,1981ApJ...244..756Y,1984A&A...132..168C,1986A&A...166...36K,1986ApJ...301..503P}. 
Generally, this microlensing results in a myriad of microimages separated by $\sim10^{-6}$ arcseconds which are unresolved with typical observing techniques. However, as the relative positions of the lens and source changes, the brightnesses of the individual microlenses change, resulting in brightness variations of the composite sources, as well as slight astrometric shifts \citep[e.g.][]{1998ApJ...501..478L,2004A&A...416...19T}. Importantly, for the purposes of this paper, this gravitational microlensing also induces time delays between microimages, but as these are  about $\sim 10^{-5}-10^{-6}$ seconds, they are typically neglected in studies of microlensed quasars \citep[see][]{2010GReGr..42.2127S}. 
However, microlensing has the potential to imprint these time delays on rapidly varying point-like sources, in particular gamma ray bursts
\citep[e.g.][]{1987ApJ...317L..51P,1997MNRAS.286L..11W,2000MNRAS.319.1163W} and Fast Radio Bursts (FRBs) \citep[e.g.][]{2016PhRvL.117i1301M,2019arXiv191207620K,2020arXiv200212778C}.

A key aspect of the formalisation of gravitational lensing is that it
can be cast as an example of Fermat's principle, where images in an optical system are observed at the extrema of a time  arrival surface \citep{1985A&A...143..413S,1986ApJ...310..568B}. 
With this, gravitational lensed images possess a parity with regards to a source dependent on whether they are formed at minima, maxima or saddle-points of the time arrival surface. 
In this paper, we will consider the impact of this macroimage parity on the microimage time delays, exploring the time arrival signature on a burst of emission from a compact source. 
Note that we will not consider diffraction and interference effects that could be apparent in the microlensing of compact sources~\citep[e.g.][]{2019arXiv191207620K,2020arXiv200201570J}, as the required time delays between microimages are significantly shorter than those being examined.

The layout of this paper is as follows; in Section~\ref{approach}, we present background on gravitational microlensing, whilst Section~\ref{parity} discusses the impact of the parity of the macroimage on the resulting microlensing time delays between images. The adopted computational approach and numerical results are presented in    Section~\ref{results}. 
The results of this paper are discussed in Section~\ref{discussion}, with the conclusions presented in Section~\ref{conclusions}. 

\section{Background and Approach}\label{approach}

\subsection{Mathematics of Microlensing}\label{time delays}
Here we consider microlensing at high optical depth, with a population of microlensing objects within the overall mass distribution of a macrolensing object. The local surface density at a macroimage is represented by a dimensionless surface mass density, $\kappa$, which is composed of a smooth mass component, $\kappa_s$, and compact mass component, $\kappa_*$, with the influence of the large-scale mass of the macrolens represented by a dimensionless shear, $\gamma$ \citep[see][for more detail]{1992grle.book.....S}. The surface density is expressed in terms of a critical surface density given by
\begin{equation}
    \Sigma_{cr} = \frac{c^2}{4 \pi G} \frac{D_{os}}{D_{ol} D_{ls}}
    \label{critical}
\end{equation}
where $D_{ij}$ are the angular diameter distances between the observer $(o)$, lens $(l)$, and source $(s)$.

\begin{figure*}
    \centering
    \includegraphics[width=2\columnwidth]{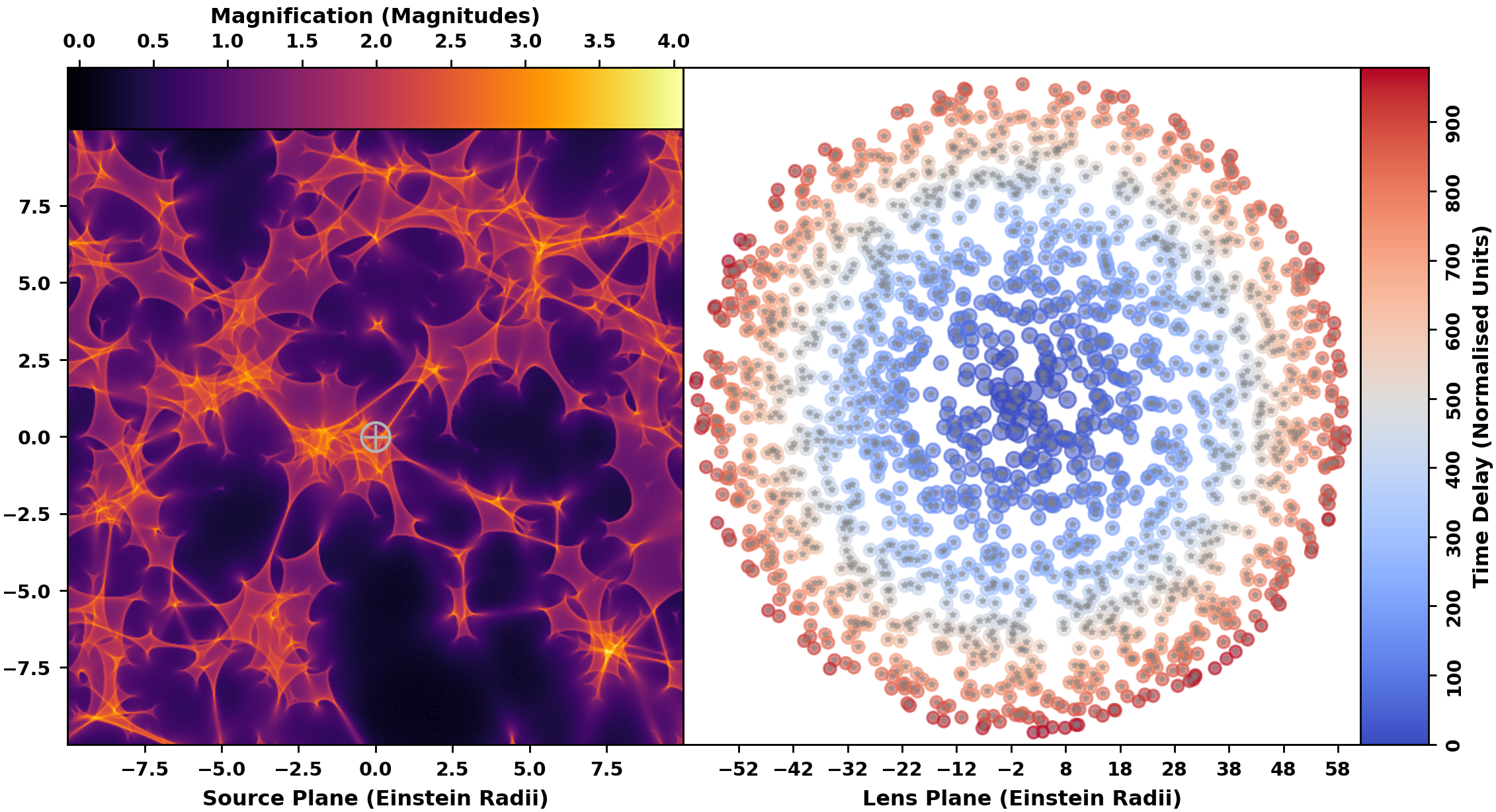}
    \caption{The results of the computational approach outlined in Section~\ref{numerical} for the case where $\kappa_* = 0.5$, $\kappa_s = 0$ and $\gamma = 0$ which corresponds two two positive eigen vectors of the magnification of the macroimage and so represents a minimum of the time arrival surface. The left-hand panel presented the map of the magnification over the source plane with the $\oplus$ symbol denoting the location of the source under consideration. The right-hand panel presents the location of the lensing masses in the lens plane, denoted as grey stars, whereas the coloured filled circles corresponds to the location of the microimages for the source location noted in the left-hand panel. Each circle is coloured with time delay, in normalised units (see Equation~\ref{timescale}), relative to the minimum of the time arrival surface. 
    }
    \label{fig1}
\end{figure*}

\begin{figure}
    \centering
    \includegraphics[width=\columnwidth]{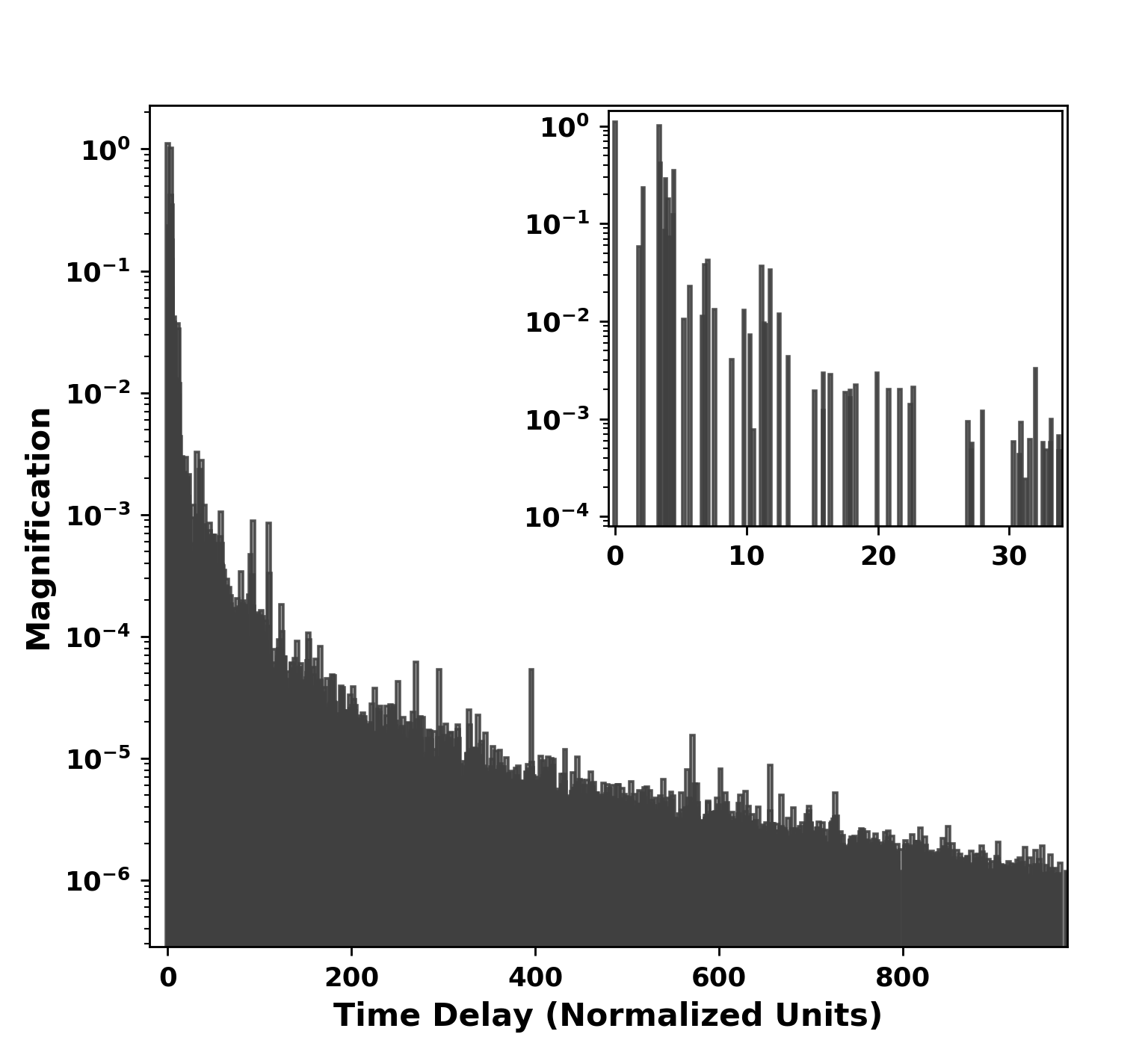}
    \caption{The magnification of the microimages as a function of time for the minimum of the time arrival of the macroimage displayed in Figure~\ref{fig1}. The inset box shows a zoom-in of the region around the brightest images. 
    }
    \label{fig1a}
\end{figure}

For a smoothly distributed lensing mass, a macroimage is formed at a location $(x_o,y_o)$, corresponding to an extremal point in the time arrival surface with value $\Phi_o$, and is mapped to a source position $(x_{so},y_{so})$. In replacing some (or all) of the smooth mass component with discrete microlensing masses, we will consider lens plane coordinates, $(x,y)$, and source plane coordinates, $(x_s,y_s)$ relative to these macroimage locations.
In the presence of a population of microlensing masses of mass, $m_i$, located at positions $(x_i,y_i)$, the local mapping between the lens and source plane coordinates is given by
\begin{equation}
    \begin{pmatrix} x_s \\ y_s \end{pmatrix}  = 
    \begin{pmatrix} 1 - \kappa_s - \gamma & 0 \\ 
                    0 & 1 - \kappa_s + \gamma 
    \end{pmatrix}
    \begin{pmatrix} x \\ y \end{pmatrix} 
    - \sum_i \frac{m_i}{r_i^{3}}
    \begin{pmatrix} x-x_i \\ y-y_i \end{pmatrix} 
    \label{mapping}
\end{equation}
where
\begin{equation}
    r_i^2 =  (x-x_i)^2 + (y-y_i)^2 
    \label{radius}
\end{equation}
and the coordinates have been oriented with regards to the direction of the external shearing.
Note that these coordinates are in units of the angular Einstein radius projected into lens and source planes, and are given by
\begin{equation}
    \xi_o = \sqrt{ \frac{4GM_\odot}{c^2}\frac{D_{ol}D_{ls}}{D_{os}}  }\ \ \ \ \ \ \ \ 
    \eta_o = \sqrt{ \frac{4GM_\odot}{c^2}\frac{D_{os}D_{ls}}{D_{ol}}  }\ \ \ resp.
    \label{einsteinradius}
\end{equation}
and the masses of the microleneses, $m_i$, are expressed in units of a solar mass.

\begin{figure*}
    \centering
    \includegraphics[width=2\columnwidth]{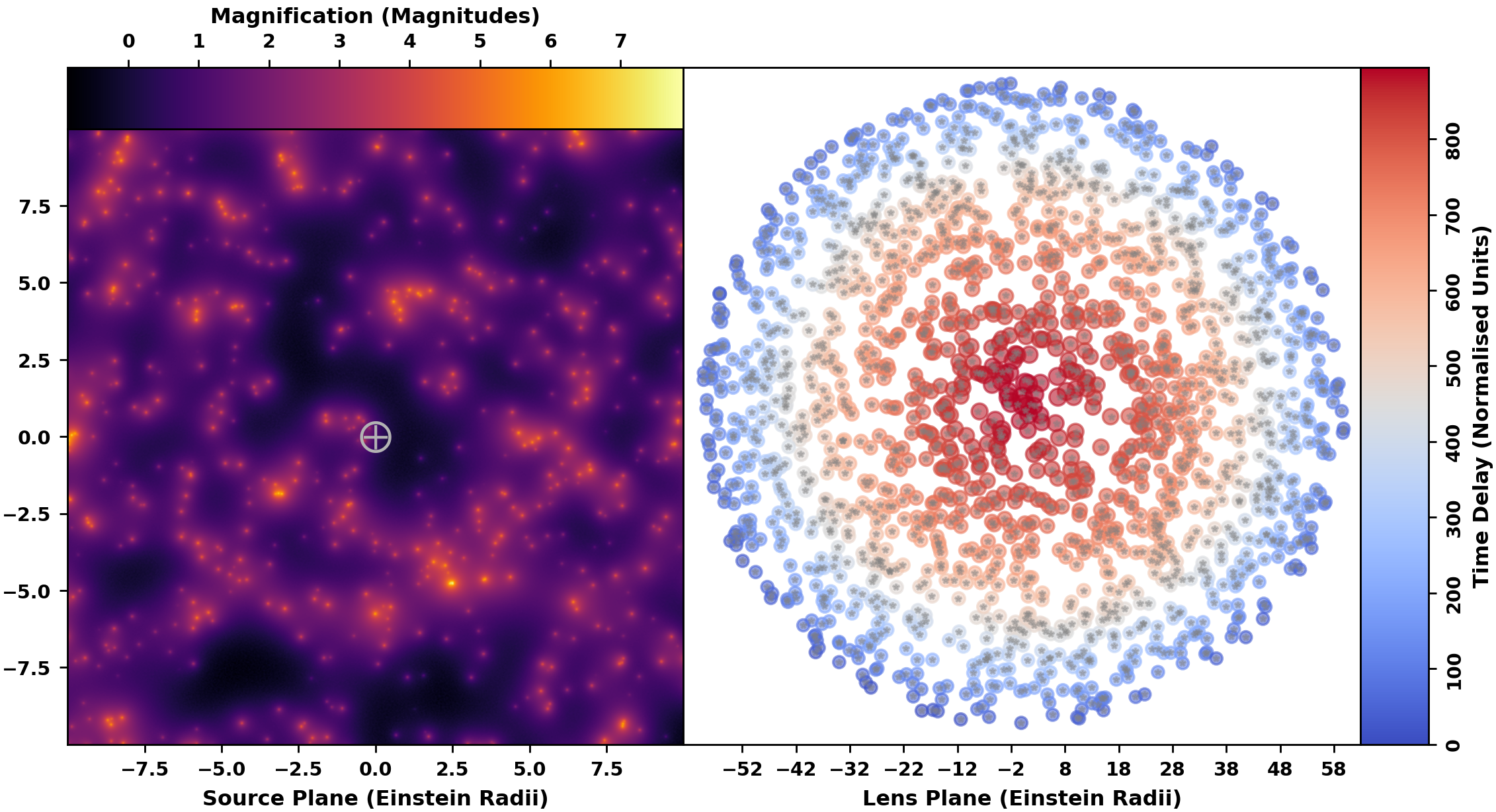}
    \caption{As Figure~\ref{fig1}, but for the case where $\kappa_* = 0.5$, $\kappa_s = 1.0$ and $\gamma = 0$, which corresponds to two negative eigen values of the macroimage magnification, representing a maximum of the time arrival surface.  
    }
    \label{fig2}
\end{figure*}

In Taylor expanding the time arrival surface about the macroimage location, we find that \citep[e.g.][]{1986ApJ...301..503P,2020MNRAS.492.1102N}; 
\begin{align}
\Phi(x,y)  = & \Phi_o - (x_o - x_{so} ) x_s - (y_o - y_{so}) y_s + \nonumber \\
 & \frac{1}{2}
\Big(
(x-x_s)^2 + (y-y_s)^2 
  - \left( \kappa_s + \gamma \right) x^2  
  - \left( \kappa_s - \gamma \right) y^2   \nonumber \\
& - \sum_i m_i \ln | r_i^2 |
\Big)
\label{time delay}    
\end{align}
This dimensionless time delay surface is transformed into physical units by a factor of
\begin{equation}
    t_o = (1+z_l) \frac{4 G M_\odot}{ c^3 }
    \label{timescale}
\end{equation}
where $z_l$ is the redshift of the lensing galaxy, corresponding to a factor of
$t_o \sim  1.97\times10^{-5} (1+z_l)$ seconds.

\begin{figure}
    \centering
    \includegraphics[width=\columnwidth]{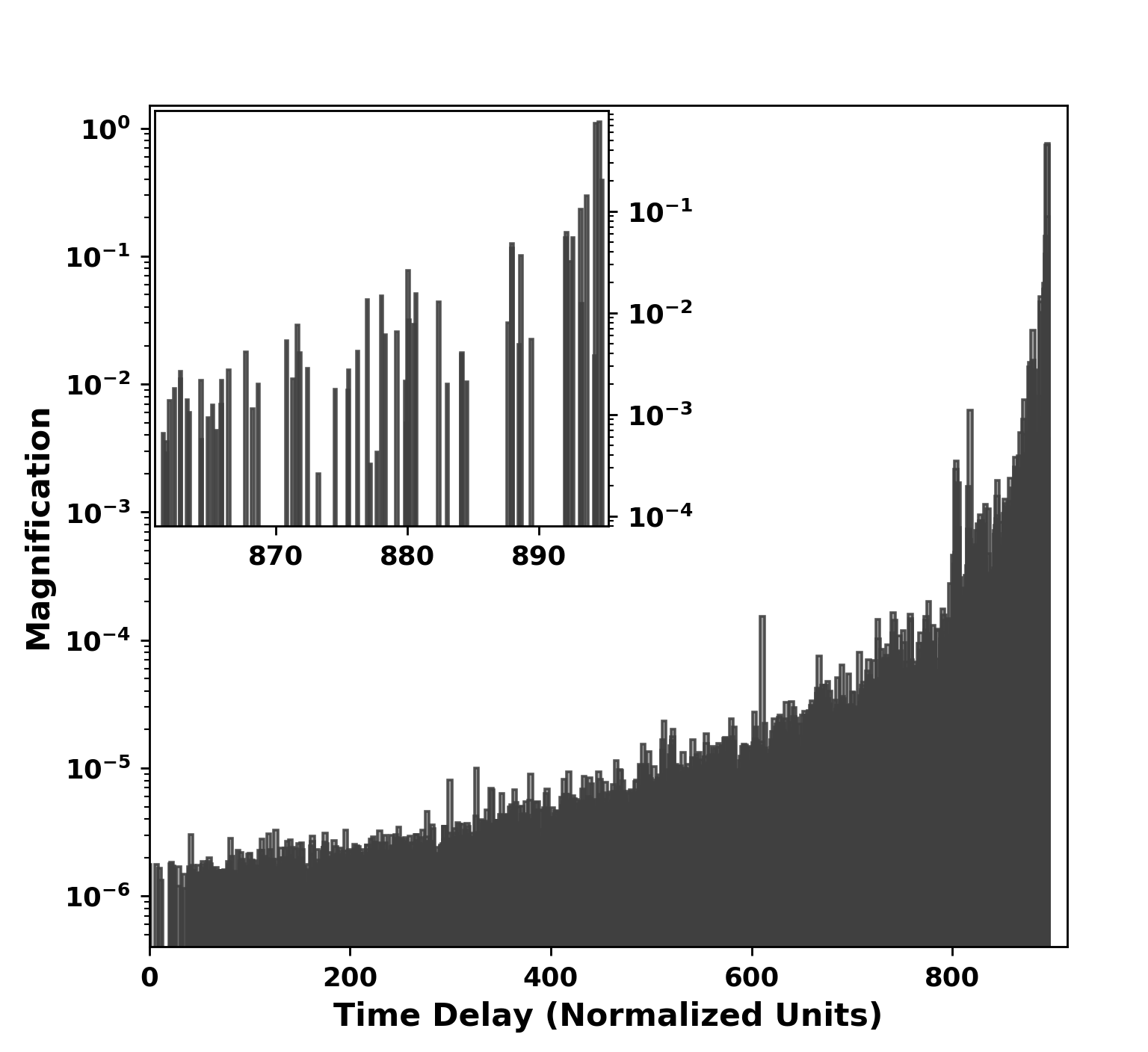}
    \caption{As Figure~\ref{fig1a}, but now considering the maximum of the macroimage time arrival surface displayed in Figure~\ref{fig2}.
    The zoom-in presents the the final moments around the peak, again separating the signal into a series of distinct peaks increasing in magnification to the final peak.
    }
    \label{fig2a}
\end{figure}

\subsection{Image Parity and Microlensing}\label{parity}
In a gravitational lens system, a small circular source is mapped to an elliptical image whose axes are scaled by the inverse of the parity eigen values of the magnification tensor \citep[again, see][for more detail]{1986ApJ...310..568B}. These two eigen values are given by;
\begin{equation}
    e_1 = ( 1 - \kappa - \gamma ) \ \ \ \ \ \ and \ \ \ \ \ \ e_2 = ( 1 - \kappa + \gamma )\ \ ,
    \label{eigen}
\end{equation}
where $\kappa$ is the total normalized surface density.
At minima of the time arrival surface, both of these eigen values are positive, whilst at maxima both eigen values are negative; 
these represent images with total positive parity.
At saddle-points of the lensing time arrival surface, one eigen value is positive and the other is negative;
these represent images with negative total parity. Locations where either of these eigen values are equal to zero denotes the location of critical lines in the image plane, with corresponding caustics in the source plane. 

\begin{figure*}
    \centering
    \includegraphics[width=2\columnwidth]{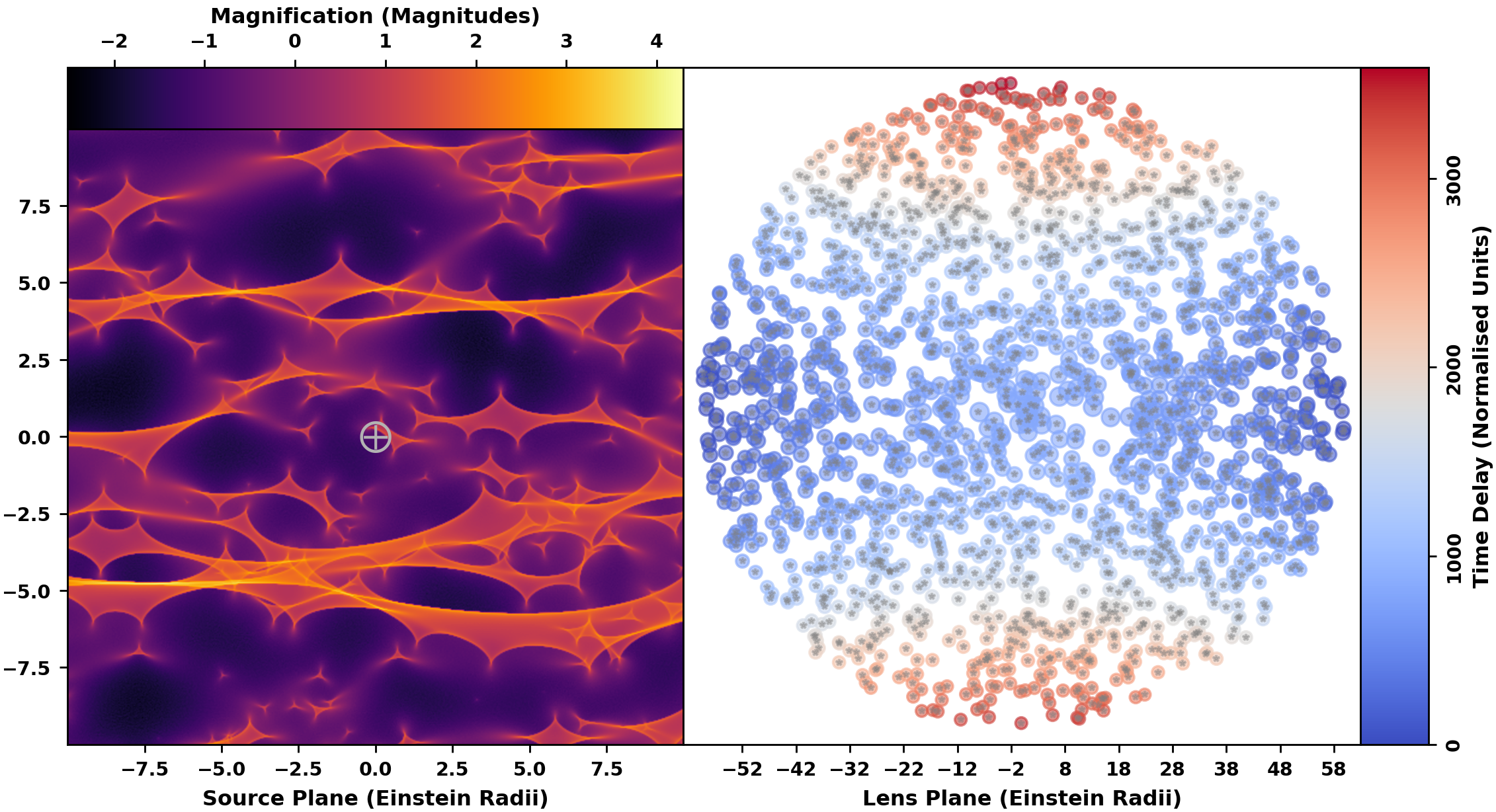}
    \caption{As Figure~\ref{fig1}, but for the case where $\kappa_* = 0.5$, $\kappa_s = 0$ and $\gamma = 1.0$, which corresponds to one positive and one negative eigen values of the macroimage magnification, representing a saddle-point of the time arrival surface.  }
    \label{fig3}
\end{figure*}

If we consider a galactic-scale lens comprised of only smooth matter, the relationship between a small source and corresponding image is described by these eigen values, a reflection of the topology of the time arrival surface.
To understand the impact of microlensing upon the time arrival surface an examination of the final term of Eqn.~\ref{time delay} reveals that each individual microlensing mass introduces a local distortion of the surface that goes as the negative logarithm of the distance to the lens. These local distortions introduce further extremal points into the time arrival surface, the locations of the myriad of microimages produced by microlensing masses. These local distortions are superimposed on the larger scale time arrival surface, enhancing the time delay between individual images. However, the magnification of these microimages falls rapidly, as $r^{-4}$, with those appearing near the location of the macroimage in the case where the mass is completely smooth appearing bright, surrounded by a halo of fainter images~\citep{1986ApJ...306....2K}.

\begin{figure}
    \centering
    \includegraphics[width=\columnwidth]{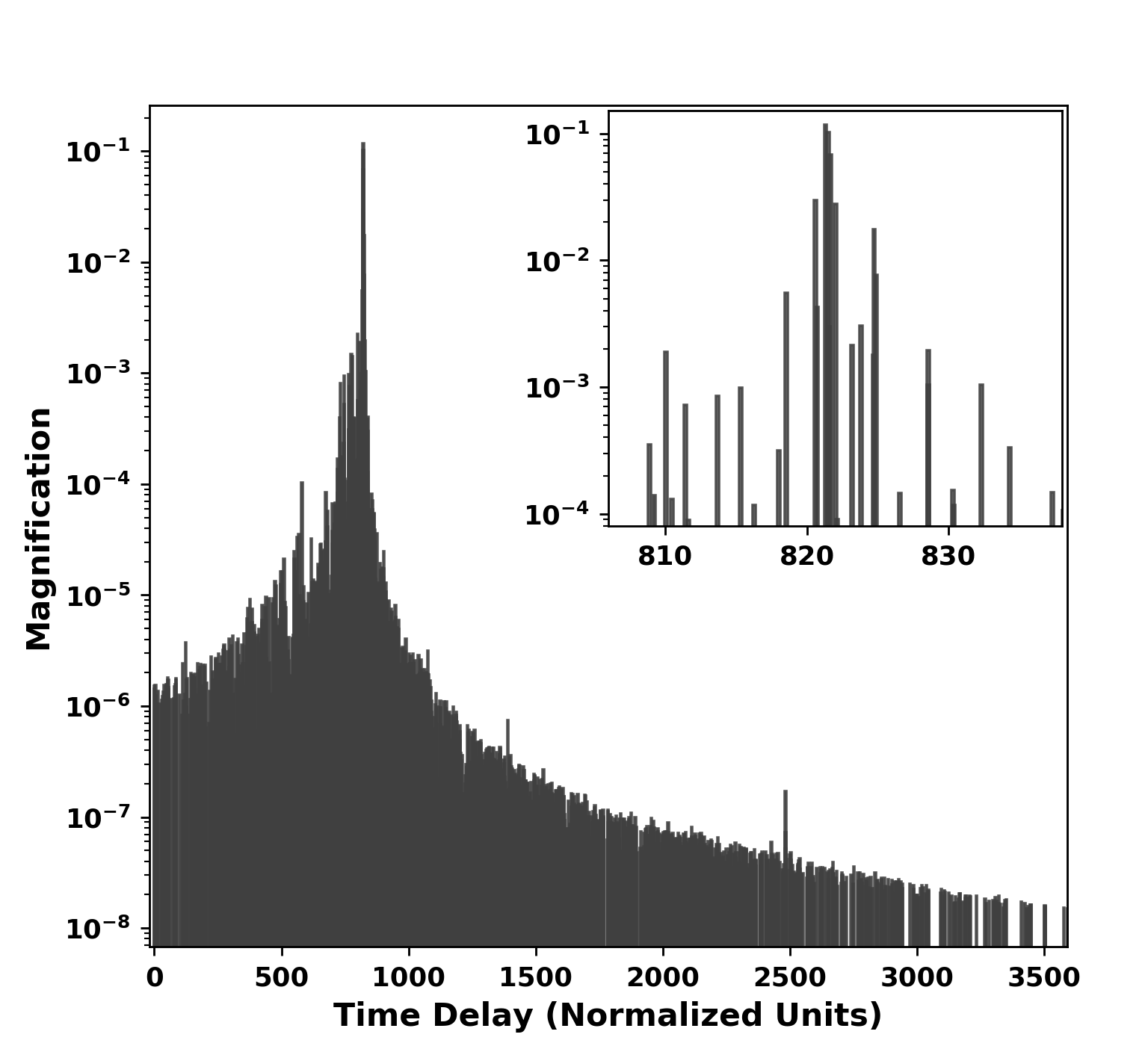}
    \caption{As Figure~\ref{fig1a}, but now considering the saddle-point of the macroimage time arrival surface displayed in Figure~\ref{fig3}.
    }
    \label{fig3a}
\end{figure}

For the purposes of this study, we will consider a single "bursty" point-like source, with a delta-function emission, reminiscent of the cosmologically compact emission from an FRB. 
Considering the topology of the time arrival surface described above, we can make some general predictions of the expected signature of the time delay between the observed images of this burst-like source. For the three potential macroimage time arrival topologies considered above:
\begin{itemize}
    \item {\it Minimum:} Remembering the picture of a few bright images surrounded by a halo of fainter images, in this situation we can expect that the burst of emission is seen firstly in the central bright image, and then, as time advances, is seen progressing through the fainter images as the signal moves outwards and climbs the time arrival surface; this behaviour can be seen in Figure~1 of \citet{1997MNRAS.286L..11W}.
    \item {\it Maximum:} This will exhibit the reverse behaviour of the minimum point, with the signal of the burst first appearing in distant faint images, then moving to progressively brighter images as the signal climbs the time arrival surface, with the appearance in the brightest microimages, at the centre of the image distribution, at the latest times.  
    \item {\it Saddle-Point:} Clearly, this represents a point intermediary between the two above cases. Here, the signal will appear in distant images along one axis, climbing out of the topological valleys before arriving that the bright images located at the saddle-point, before moving again to fainter images as the signal ascends the sides out of the saddle-point towards the topological peaks. 
    We note that \citet{1997MNRAS.286L..11W} did consider a case with of an image configuration representing a saddle point of the time arrival surface, they did not explicitly examine the form of the time delay signal of the microimages.
\end{itemize}
To explore this further, in the following we will consider the influence of gravitational microlensing on the time delay surface and the resulting impact on the temporal signature of a bursting source for several fiducial models, with a more general exploration presented in a future publication.

\subsection{Numerical Approach}\label{numerical}
As with previous studies of the gravitational microlensing of quasars, we employ backwards ray tracing to calculate a map of the microlensing magnification. The details of this approach has been presented in the literature \citep[see][for a detailed discussion]{1990LNP...360..186W} and only a brief summary is presented here. 

To numerically explore the impact of image parity on the microlensing time delay signal, equations~\ref{mapping} and \ref{time delay} were implements in {\tt python}.
Given that the mapping in Equation~\ref{mapping} is one-to-many, analytically locating the positions of images for a particular source location and a distribution of lensing masses is not tractable, and so a series of stages were developed. Firstly, a distribution of $n\times m$ rays are traced through the lens plane and collected in the source \citep[equivalent to the backwards ray-tracing approach of ][]{1990LNP...360..186W}, with the ray-tracing efficiency improved through the implementation of {\tt ipyparallel} and multiple CPUs. For each ray, the lens and corresponding source plane locations were recorded, as well as the time delay and magnification, allowing for the construction of a time delay map over the lens plane, and magnification maps for both the source and lens plane. 

As we are consider a point-like source, it is extremely unlikely that a ray traced through the image plane lands precisely in a source location. Given this, rays that land within a small region of the source are used as a starting point for a minimisation approach that adjusts the ray location in the source plane so that it coincides with the source. Furthermore, as we expect that there will be an image associated with each star, a similar minimisation search is undertaken at a set of random set of points in the vicinity of each star location. These minimisation searches locate the positions of the images, but given the number of random search starts, it is likely that some images are located multiple times, and so the solutions are sorted and multiple entries are removed. As pointed out by \citet{1987ApJ...317L..51P}, we note that any approach is not guaranteed to locate every image, but a sufficiently large exploration should locate the vast majority of images, especially the significantly bright images.

\section{Results}\label{results}
In the following, we will consider the topologies of the time arrival surface as discussed in Section~\ref{parity} through selection of specific combinations of microlensing parameters. In each of the numerical explorations, a source plane region of 20$\times$20 Einstein radii was considered, to obtain a magnification map over the source plane, rays traced through a corresponding region of the lens plane to determine the locations strongly magnified images. An additional buffering region in terms of the considered stars in added to consider the impact of images of the source observed far from the bright few images; whilst we expect these additional images to be demagnified with regards to the source, this will reveal the dependence upon the topology of the time arrival surface. The stars are distributed randomly over the lens plane and all possess a mass of $1 M_\odot$. 
The three cases considered in this paper:

\subsection{Minimum}\label{minimum}
The first case  has $\kappa_* = 0.5$, with $\kappa_s$ and $\gamma$ equal to zero, which results in two positive parity eigen vectors (Equation~\ref{eigen}), so this configuration represents a minimum of the time arrival surface. The numerical results for this configuration is presented in Figure~\ref{fig1}, with the left-hand panel presents the magnification map \citep[see][for details]{1990LNP...360..186W} for the particular distribution of lensing masses presented in the right-hand panel as grey stars; as expected, a rich caustic network is apparent over the source plane. 
Additionally, the right-hand panel presents the locations of the images are identified through the numerical search outlined above; as expected, at least one image is identified for each of the microlensing stars. Each of the images is colour-coded with the time delay between the images, relative to the first that responses burst of the source. Quite clearly, this reflects the behaviour discussed in Section~\ref{parity}, with the central images lighting up first, with the signal spreading outwards to fainter images at larger radius.

To explore this in more detail, we consider the time delay between the microimages verses their magnification, presenting the resulting distribution in Figure~\ref{fig1a}. As expected, the signal begins with images with magnification of order unity before progressing to fainter images, although noting that the magnification in this figure is presented on a logarithmic scale. To further emphasise the behaviour, the inset box zooms in on the initial response to the burst in the source, resolving the response signal into a series of individual spikes over the period, again showing the behaviour of the initial images showing the largest magnification, followed by images with smaller magnification, but displaying some stochasticity in their response. 

\subsection{Maximum}\label{maximum}
To represent a maximum of the time arrival surface, we considered it situation with microlensing parameters of $\kappa_* = 0.5$ and $\kappa_s = 1.0$, with $\gamma = 0$. With these, both of the parity eigen values are negative. The results of the numerical exploration is presented in Figure~\ref{fig2}, which has the same format as Figure~\ref{fig1}. It is instructive to note that the caustic network with such a significant component of smooth matter is substantially different to the case with only compact lenses. 

Examining the right-hand panel of Figure~\ref{fig2} reveals the opposite signal to that presented in Section~\ref{minimum}. Here the signal appears firstly in the distant, fainter sources, before moving to smaller radii and brighter images. Hence, the underlying macroimage time arrival surface is imprinted on the microlensing time delay signature. To explore this further, we again considered the time delay verses image brightness distribution, presenting the distribution in Figure~\ref{fig2a}. As expected, the signal begins with faint images and increases towards the brightest images, with the zoom-in displaying this behaviour with the discrete emission spikes.

\subsection{Saddle-Point}\label{saddle}
The final situation explored in this paper considered $\kappa_* = 0.5$, with $\kappa_s = 0$ and $\gamma = 1.0$; with this, there is one positive parity eigen value, whilst the other is negative, demonstrating that the time arrival surface is a saddle point. The results are presented in Figure~\ref{fig3}, with the left-hand panel displaying the stretched caustic network due to the significant shearing \citep[again see][for a more detailed exploration]{1990LNP...360..186W}.

The image time delays presented in the right-hand panel of Figure~\ref{fig3} demonstrates the behaviour outlined in Section~\ref{parity}, with distant, faint images along one access lighting up first from the source burst, before moving towards the central, brighter images, and then again out to fainter images along the perpendicular axis. 
Again, we can consider the brightness of the microimages as a function of arrival time, as presented in Figure~\ref{fig3a} which confirms this behaviour. It is noticeable that the resultant signal of time delays for the microimages does not possess the sharp boundary seen in the maximal or minimal case. However, it is clear that the distribution is asymmetric declining more rapidly than it increases; this is something we explore in more detail below. Again, this behaviour is reflected in the zoom-in, with the distribution increasing to a peak and then declining.

\section{Discussion}\label{discussion} 

\subsection{Observed Signatures}\label{observedsignature}
This paper has consider gravitational microlensing at high optical depths, with significant surface densities and/or shearing, 
 and hence will be encountered when a source is multiply imaged my a galactic scale mass;  the physics of this scenario is well understood with numerous examples of multiply imaged quasars and galaxies identified~\citep[e.g.][]{2010ARA&A..48...87T}. Hence, a source such as an FRB appropriately aligned with a galactic scale mass will be multiply imaged. These images will be localised on the sky, on the scale of arcseconds, and will appear with hours to months time delays between each of these macroimages~\citep{2016A&ARv..24...11T}. 

Examining the topology of the time arrival surface depends upon the properties of the macrolensing mass distribution,~\citep{1985A&A...143..413S,1986ApJ...310..568B}, but in general the first image will occur at a minima, progressing through a series of other minima, saddle-points or maxima, to the final image appearing at a maximum of the time arrival surface. Note that if the lensing mass is singular, this final maximum of the time arrival surface is also singular, and so the resultant image will be infinitely demagnified and unobservable~\citep[see][]{1992grle.book.....S}.   
Each of the macroimages will be susceptible to microlensing, dependent upon the local mass density and shearing. Hence, the microlensing of the first image to appear will imprint a temporal signature as outlined in Section~\ref{minimum}, beginning with bright images and fading through fainter images. 
The form of the microlensing of the subsequent images will be imprinted with the temporal signature dependent upon the local topology of the time arrival surface, with the final image, if observable, will occur at a maximum and so will build up from faint microimages to finish with bight microimages (Section~\ref{maximum}).

It has already been recognised that short duration of FRBs means that they represent ideal cosmological probes through the measurement of gravitational lensing time delays~\citep[e.g.][]{2018NatCo...9.3833L,2019A&A...621A..91W,2019PhRvD..99l3517L}, with additional constraints coming from localisation of optical afterglows~\citep[e.g.][]{2020arXiv200513159M}. However, the biggest source of uncertainty in gravitational lensing reconstruction comes from the determination of the underlying mass distribution~\citep[see extensive work by the H0LiCOW program e.g.][and subsequent publications]{2017MNRAS.468.2590S}.
The identification of the temporal signatures imprinted by gravitational microlensing will reveal the underlying macroimage parities, providing additional constraints in determining the underlying galactic mass distribution, reducing the uncertainty. In a subsequent contribution, we will explore the dependence of the temporal signature on the macro- and microlensing parameters, and whether the precise form of the microlensing temporal signature will provide additional constraints on overall gravitational lensing model.

\subsection{Observability}\label{observability}
It is clear that FRBs are at cosmological distances, now revealing the baryonic content of the universe on gigaparsec scales~\citep{2020Natur.581..391M}. Assuming a typical redshift for a gravitational lens of $z_l=0.4$, the corresponding physical time units is (Eqn.~\ref{timescale}) of $t_o \sim 0.03 ms$, and the duration of the inset boxes of Figures~\ref{fig1a}, \ref{fig2a} and \ref{fig3a} corresponds to $\sim 1 ms$, accessible to modern instruments such as the Australian Square Kilometre Array Pathfinder (ASKAP) with a resolution of 10s of $\mu s$ \citep[e.g.][]{2018MNRAS.478.1209F,2020arXiv200513162D}. Furthermore, the individual peaks of FRBs can occur on substantially less than $ms$ time scales, indicating that the individual microimages can be temporally resolvable~\citep{2020ApJ...891L..38C}. 
Whilst the relative brightening/fading of the temporal signature is rapid over this period, ranging from magnifications ranging from around unity to a factor of $\sim 10^{-3}$ over this period. However, there is a growing number of FRBs that have been observed with $S/N$ exceeding 100~\citep[see][and {\it frbcat.org}]{2016PASA...33...45P}, and hence, for a sufficiently bright FRB observed through a gravitational lens, the signature of individual microimages should be apparent over the $ms$ time scales, even if demagnified by a factor of more than a fact of $\sim 100$.  

\section{Conclusions}\label{conclusions}
This paper has examined the impact of image parity on the temporal signature due to the action of gravitational microlensing of FRB sources. Three image configurations, defined by the signs of the eigen values of the magnification matrix, were considered, corresponding respectively to a minimum, a maximum and a saddle-point of the macroimage time arrival surface. The key finding is that the topology of the time arrival surface is imprinted on the time delay between microimages. If a variable point-like source, in particularly one exhibiting bursts, is seen through these microlensing masses, their observed variability will be modulated by this signal, with each of the macroimage topologies imprinting distinct pattern on the resultant light curve.      

Further, we considered the impact of microlensing on FRBs, finding their short duration, and observational cadence available with current and future radio facilities, make them ideal targets to search for the impact of gravitational microlensing. Given that the high optical depths considered in this paper occur in situations where multiple imaging my a galactic scale lens, the FRBs will be expected to repeat on the hours to years time scales of strong lensing, with the microlensing signature imprinted on each distinct macrolensed image. However, given the overall topology of the time arrival surface, there will be a particular ordering beginning with a minimum and finishing with a maximum, although this final image could be highly demagnified if the lensing mass is singular. The resolution of optical afterglows, to pin-point the macrolens image locations, coupled with the additional information provided by image parities, has the potential to add to the ability of FRBs to constrain cosmological models.  

The focus of this paper has been three specific examples of microlensing with appropriate topologies of the time arrival surface to demonstrate its impact. 
A more extensive exploration of the impact of the time arrival topologies on the microlensing signature of FRBs, focused upon key variables, including the mass function of microleses, smooth matter content, and the location of the source relative to the caustic network, will be presented in a future publication. 

\section*{Acknowledgements}
GFL thanks the anonymous referee, Tara Murphy for useful insights on the nature of FRBs, Angela Ng for comments on the paper, and  Mark Walker for the invitation to speak at the Manly Astrophysics meeting on "Lensing Of Fast Transients", where the seeds of this study were (re-)sewn.
GFL would also like to thank Andrew Janke of the University of Sydney's Information and Communication Technologies (ICT) for provisioning access to significant cloud computational resources on a Saturday morning, ensuring the smooth(ish) running of this project through the corona virus pandemic.
Furthermore, GFL acknowledges computational support through the Sydney Informatics Hub, a Research Core Facility (RCF) at the University of Sydney.
This research has made use of NASA’s Astrophysics Data System, {\tt matplotlib} \citep{Hunter:2007}, {\tt scipy/numpy} \citep{2020SciPy-NMeth} and {\tt ipyparallel} ({\tt ipyparallel.readthedocs.io}).

\section*{Data Availability}
The data generated as part of this project may be shared on a reasonable request to the corresponding author.



\bibliographystyle{mnras}
\bibliography{MicrolensingPaper} 

\begin{thebibliography}{}
\makeatletter
\relax
\def\mn@urlcharsother{\let\do\@makeother \do\$\do\&\do\#\do\^\do\_\do\%\do\~}
\def\mn@doi{\begingroup\mn@urlcharsother \@ifnextchar [ {\mn@doi@}
  {\mn@doi@[]}}
\def\mn@doi@[#1]#2{\def\@tempa{#1}\ifx\@tempa\@empty \href
  {http://dx.doi.org/#2} {doi:#2}\else \href {http://dx.doi.org/#2} {#1}\fi
  \endgroup}
\def\mn@eprint#1#2{\mn@eprint@#1:#2::\@nil}
\def\mn@eprint@arXiv#1{\href {http://arxiv.org/abs/#1} {{\tt arXiv:#1}}}
\def\mn@eprint@dblp#1{\href {http://dblp.uni-trier.de/rec/bibtex/#1.xml}
  {dblp:#1}}
\def\mn@eprint@#1:#2:#3:#4\@nil{\def\@tempa {#1}\def\@tempb {#2}\def\@tempc
  {#3}\ifx \@tempc \@empty \let \@tempc \@tempb \let \@tempb \@tempa \fi \ifx
  \@tempb \@empty \def\@tempb {arXiv}\fi \@ifundefined
  {mn@eprint@\@tempb}{\@tempb:\@tempc}{\expandafter \expandafter \csname
  mn@eprint@\@tempb\endcsname \expandafter{\@tempc}}}

\bibitem[\protect\citeauthoryear{{Blandford} \& {Narayan}}{{Blandford} \&
  {Narayan}}{1986}]{1986ApJ...310..568B}
{Blandford} R.,  {Narayan} R.,  1986, \mn@doi [\apj] {10.1086/164709}, \href
  {https://ui.adsabs.harvard.edu/abs/1986ApJ...310..568B} {310, 568}

\bibitem[\protect\citeauthoryear{{Carr}, {Kohri}, {Sendouda}  \&
  {Yokoyama}}{{Carr} et~al.}{2020}]{2020arXiv200212778C}
{Carr} B.,  {Kohri} K.,  {Sendouda} Y.,   {Yokoyama} J.,  2020, arXiv e-prints,
  \href {https://ui.adsabs.harvard.edu/abs/2020arXiv200212778C} {p.
  arXiv:2002.12778}

\bibitem[\protect\citeauthoryear{{Chang} \& {Refsdal}}{{Chang} \&
  {Refsdal}}{1979}]{1979Natur.282..561C}
{Chang} K.,  {Refsdal} S.,  1979, \mn@doi [\nat] {10.1038/282561a0}, \href
  {https://ui.adsabs.harvard.edu/abs/1979Natur.282..561C} {282, 561}

\bibitem[\protect\citeauthoryear{{Chang} \& {Refsdal}}{{Chang} \&
  {Refsdal}}{1984}]{1984A&A...132..168C}
{Chang} K.,  {Refsdal} S.,  1984, \aap, \href
  {https://ui.adsabs.harvard.edu/abs/1984A&A...132..168C} {132, 168}

\bibitem[\protect\citeauthoryear{{Cho} et~al.,}{{Cho}
  et~al.}{2020}]{2020ApJ...891L..38C}
{Cho} H.,  et~al., 2020, \mn@doi [\apjl] {10.3847/2041-8213/ab7824}, \href
  {https://ui.adsabs.harvard.edu/abs/2020ApJ...891L..38C} {891, L38}

\bibitem[\protect\citeauthoryear{{Day} et~al.,}{{Day}
  et~al.}{2020}]{2020arXiv200513162D}
{Day} C.~K.,  et~al., 2020, arXiv e-prints, \href
  {https://ui.adsabs.harvard.edu/abs/2020arXiv200513162D} {p. arXiv:2005.13162}

\bibitem[\protect\citeauthoryear{{Farah} et~al.,}{{Farah}
  et~al.}{2018}]{2018MNRAS.478.1209F}
{Farah} W.,  et~al., 2018, \mn@doi [\mnras] {10.1093/mnras/sty1122}, \href
  {https://ui.adsabs.harvard.edu/abs/2018MNRAS.478.1209F} {478, 1209}

\bibitem[\protect\citeauthoryear{Hunter}{Hunter}{2007}]{Hunter:2007}
Hunter J.~D.,  2007, \mn@doi [Computing in Science \& Engineering]
  {10.1109/MCSE.2007.55}, 9, 90

\bibitem[\protect\citeauthoryear{{Jow}, {Foreman}, {Pen}  \& {Zhu}}{{Jow}
  et~al.}{2020}]{2020arXiv200201570J}
{Jow} D.~L.,  {Foreman} S.,  {Pen} U.-L.,   {Zhu} W.,  2020, arXiv e-prints,
  \href {https://ui.adsabs.harvard.edu/abs/2020arXiv200201570J} {p.
  arXiv:2002.01570}

\bibitem[\protect\citeauthoryear{{Katz}, {Balbus}  \& {Paczynski}}{{Katz}
  et~al.}{1986}]{1986ApJ...306....2K}
{Katz} N.,  {Balbus} S.,   {Paczynski} B.,  1986, \mn@doi [\apj]
  {10.1086/164313}, \href
  {https://ui.adsabs.harvard.edu/abs/1986ApJ...306....2K} {306, 2}

\bibitem[\protect\citeauthoryear{{Katz}, {Kopp}, {Sibiryakov}  \& {Xue}}{{Katz}
  et~al.}{2019}]{2019arXiv191207620K}
{Katz} A.,  {Kopp} J.,  {Sibiryakov} S.,   {Xue} W.,  2019, arXiv e-prints,
  \href {https://ui.adsabs.harvard.edu/abs/2019arXiv191207620K} {p.
  arXiv:1912.07620}

\bibitem[\protect\citeauthoryear{{Kayser}, {Refsdal}  \& {Stabell}}{{Kayser}
  et~al.}{1986}]{1986A&A...166...36K}
{Kayser} R.,  {Refsdal} S.,   {Stabell} R.,  1986, \aap, \href
  {https://ui.adsabs.harvard.edu/abs/1986A&A...166...36K} {166, 36}

\bibitem[\protect\citeauthoryear{{Lewis} \& {Ibata}}{{Lewis} \&
  {Ibata}}{1998}]{1998ApJ...501..478L}
{Lewis} G.~F.,  {Ibata} R.~A.,  1998, \mn@doi [\apj] {10.1086/305860}, \href
  {https://ui.adsabs.harvard.edu/abs/1998ApJ...501..478L} {501, 478}

\bibitem[\protect\citeauthoryear{{Li}, {Gao}, {Ding}, {Wang}  \& {Zhang}}{{Li}
  et~al.}{2018}]{2018NatCo...9.3833L}
{Li} Z.-X.,  {Gao} H.,  {Ding} X.-H.,  {Wang} G.-J.,   {Zhang} B.,  2018,
  \mn@doi [Nature Communications] {10.1038/s41467-018-06303-0}, \href
  {https://ui.adsabs.harvard.edu/abs/2018NatCo...9.3833L} {9, 3833}

\bibitem[\protect\citeauthoryear{{Liu}, {Li}, {Gao}  \& {Zhu}}{{Liu}
  et~al.}{2019}]{2019PhRvD..99l3517L}
{Liu} B.,  {Li} Z.,  {Gao} H.,   {Zhu} Z.-H.,  2019, \mn@doi [\prd]
  {10.1103/PhysRevD.99.123517}, \href
  {https://ui.adsabs.harvard.edu/abs/2019PhRvD..99l3517L} {99, 123517}

\bibitem[\protect\citeauthoryear{{Macquart} et~al.,}{{Macquart}
  et~al.}{2020}]{2020Natur.581..391M}
{Macquart} J.~P.,  et~al., 2020, \mn@doi [\nat] {10.1038/s41586-020-2300-2},
  \href {https://ui.adsabs.harvard.edu/abs/2020Natur.581..391M} {581, 391}

\bibitem[\protect\citeauthoryear{{Marnoch} et~al.,}{{Marnoch}
  et~al.}{2020}]{2020arXiv200513159M}
{Marnoch} L.,  et~al., 2020, arXiv e-prints, \href
  {https://ui.adsabs.harvard.edu/abs/2020arXiv200513159M} {p. arXiv:2005.13159}

\bibitem[\protect\citeauthoryear{{Mu{\~n}oz}, {Kovetz}, {Dai}  \&
  {Kamionkowski}}{{Mu{\~n}oz} et~al.}{2016}]{2016PhRvL.117i1301M}
{Mu{\~n}oz} J.~B.,  {Kovetz} E.~D.,  {Dai} L.,   {Kamionkowski} M.,  2016,
  \mn@doi [\prl] {10.1103/PhysRevLett.117.091301}, \href
  {https://ui.adsabs.harvard.edu/abs/2016PhRvL.117i1301M} {117, 091301}

\bibitem[\protect\citeauthoryear{{Ng} \& {Lewis}}{{Ng} \&
  {Lewis}}{2020}]{2020MNRAS.492.1102N}
{Ng} A. L.~H.,  {Lewis} G.~F.,  2020, \mn@doi [\mnras] {10.1093/mnras/stz3475},
  \href {https://ui.adsabs.harvard.edu/abs/2020MNRAS.492.1102N} {492, 1102}

\bibitem[\protect\citeauthoryear{{Paczynski}}{{Paczynski}}{1986}]{1986ApJ...301..503P}
{Paczynski} B.,  1986, \mn@doi [\apj] {10.1086/163919}, \href
  {https://ui.adsabs.harvard.edu/abs/1986ApJ...301..503P} {301, 503}

\bibitem[\protect\citeauthoryear{{Paczynski}}{{Paczynski}}{1987}]{1987ApJ...317L..51P}
{Paczynski} B.,  1987, \mn@doi [\apjl] {10.1086/184911}, \href
  {https://ui.adsabs.harvard.edu/abs/1987ApJ...317L..51P} {317, L51}

\bibitem[\protect\citeauthoryear{{Petroff} et~al.,}{{Petroff}
  et~al.}{2016}]{2016PASA...33...45P}
{Petroff} E.,  et~al., 2016, \mn@doi [\pasa] {10.1017/pasa.2016.35}, \href
  {https://ui.adsabs.harvard.edu/abs/2016PASA...33...45P} {33, e045}

\bibitem[\protect\citeauthoryear{{Refsdal}}{{Refsdal}}{1964}]{1964MNRAS.128..307R}
{Refsdal} S.,  1964, \mn@doi [\mnras] {10.1093/mnras/128.4.307}, \href
  {https://ui.adsabs.harvard.edu/abs/1964MNRAS.128..307R} {128, 307}

\bibitem[\protect\citeauthoryear{{Refsdal}}{{Refsdal}}{1966}]{1966MNRAS.132..101R}
{Refsdal} S.,  1966, \mn@doi [\mnras] {10.1093/mnras/132.1.101}, \href
  {https://ui.adsabs.harvard.edu/abs/1966MNRAS.132..101R} {132, 101}

\bibitem[\protect\citeauthoryear{{Schmidt} \& {Wambsganss}}{{Schmidt} \&
  {Wambsganss}}{2010}]{2010GReGr..42.2127S}
{Schmidt} R.~W.,  {Wambsganss} J.,  2010, \mn@doi [General Relativity and
  Gravitation] {10.1007/s10714-010-0956-x}, \href
  {https://ui.adsabs.harvard.edu/abs/2010GReGr..42.2127S} {42, 2127}

\bibitem[\protect\citeauthoryear{{Schneider}}{{Schneider}}{1985}]{1985A&A...143..413S}
{Schneider} P.,  1985, \aap, \href
  {https://ui.adsabs.harvard.edu/abs/1985A&A...143..413S} {143, 413}

\bibitem[\protect\citeauthoryear{{Schneider}, {Ehlers}  \& {Falco}}{{Schneider}
  et~al.}{1992}]{1992grle.book.....S}
{Schneider} P.,  {Ehlers} J.,   {Falco} E.~E.,  1992, {Gravitational Lenses},
  \mn@doi{10.1007/978-3-662-03758-4.
}

\bibitem[\protect\citeauthoryear{{Suyu} et~al.,}{{Suyu}
  et~al.}{2017}]{2017MNRAS.468.2590S}
{Suyu} S.~H.,  et~al., 2017, \mn@doi [\mnras] {10.1093/mnras/stx483}, \href
  {https://ui.adsabs.harvard.edu/abs/2017MNRAS.468.2590S} {468, 2590}

\bibitem[\protect\citeauthoryear{{Treu}}{{Treu}}{2010}]{2010ARA&A..48...87T}
{Treu} T.,  2010, \mn@doi [\araa] {10.1146/annurev-astro-081309-130924}, \href
  {https://ui.adsabs.harvard.edu/abs/2010ARA&A..48...87T} {48, 87}

\bibitem[\protect\citeauthoryear{{Treu} \& {Marshall}}{{Treu} \&
  {Marshall}}{2016}]{2016A&ARv..24...11T}
{Treu} T.,  {Marshall} P.~J.,  2016, \mn@doi [\aapr]
  {10.1007/s00159-016-0096-8}, \href
  {https://ui.adsabs.harvard.edu/abs/2016A&ARv..24...11T} {24, 11}

\bibitem[\protect\citeauthoryear{{Treyer} \& {Wambsganss}}{{Treyer} \&
  {Wambsganss}}{2004}]{2004A&A...416...19T}
{Treyer} M.,  {Wambsganss} J.,  2004, \mn@doi [\aap]
  {10.1051/0004-6361:20034284}, \href
  {https://ui.adsabs.harvard.edu/abs/2004A&A...416...19T} {416, 19}

\bibitem[\protect\citeauthoryear{{Virtanen} et~al.,}{{Virtanen}
  et~al.}{2020}]{2020SciPy-NMeth}
{Virtanen} P.,  et~al., 2020, \mn@doi [Nature Methods]
  {https://doi.org/10.1038/s41592-019-0686-2}, \href {https://rdcu.be/b08Wh}
  {17, 261}

\bibitem[\protect\citeauthoryear{{Wagner}, {Liesenborgs}  \&
  {Eichler}}{{Wagner} et~al.}{2019}]{2019A&A...621A..91W}
{Wagner} J.,  {Liesenborgs} J.,   {Eichler} D.,  2019, \mn@doi [\aap]
  {10.1051/0004-6361/201833530}, \href
  {https://ui.adsabs.harvard.edu/abs/2019A&A...621A..91W} {621, A91}

\bibitem[\protect\citeauthoryear{{Wambsganss}}{{Wambsganss}}{1990}]{1990LNP...360..186W}
{Wambsganss} J.,  1990, {Microlensing calculations with a hierarchical tree
  code: New results}.
p.~186, \mn@doi{10.1007/BFb0009253}

\bibitem[\protect\citeauthoryear{{Williams} \& {Wijers}}{{Williams} \&
  {Wijers}}{1997}]{1997MNRAS.286L..11W}
{Williams} L.~L.~R.,  {Wijers} R.~A.~M.~J.,  1997, \mn@doi [\mnras]
  {10.1093/mnras/286.1.L11}, \href
  {https://ui.adsabs.harvard.edu/abs/1997MNRAS.286L..11W} {286, L11}

\bibitem[\protect\citeauthoryear{{Wong} et~al.,}{{Wong}
  et~al.}{2019}]{2019arXiv190704869W}
{Wong} K.~C.,  et~al., 2019, arXiv e-prints, \href
  {https://ui.adsabs.harvard.edu/abs/2019arXiv190704869W} {p. arXiv:1907.04869}

\bibitem[\protect\citeauthoryear{{Wyithe} \& {Turner}}{{Wyithe} \&
  {Turner}}{2000}]{2000MNRAS.319.1163W}
{Wyithe} J.~S.~B.,  {Turner} E.~L.,  2000, \mn@doi [\mnras]
  {10.1046/j.1365-8711.2000.03919.x}, \href
  {https://ui.adsabs.harvard.edu/abs/2000MNRAS.319.1163W} {319, 1163}

\bibitem[\protect\citeauthoryear{{Young}}{{Young}}{1981}]{1981ApJ...244..756Y}
{Young} P.,  1981, \mn@doi [\apj] {10.1086/158752}, \href
  {https://ui.adsabs.harvard.edu/abs/1981ApJ...244..756Y} {244, 756}

\bibitem[\protect\citeauthoryear{Zwicky}{Zwicky}{1937}]{PhysRev.51.290}
Zwicky F.,  1937, \mn@doi [Phys. Rev.] {10.1103/PhysRev.51.290}, 51, 290

\makeatother
\end{thebibliography}

\bsp	
\label{lastpage}
\end{document}